\documentclass[reprint,
showpacs,amsmath,amssymb,aps,prl]{revtex4-1}

\usepackage[dvipdfm]{graphicx}
\usepackage{bm}

\begin{document}

\title{Shubnikov-de Haas oscillations of a single layer graphene under dc current bias}
\author{Zhenbing Tan}
\author{ChangLing Tan}
\author{Li Ma}
\author{G. T. Liu}
\author{L. Lu}
\author{C. L. Yang}
\affiliation{Daniel Chee Tsui Laboratory, Beijing National Laboratory for Condensed Matter
Physics, Institute of Physics, Chinese Academy of Sciences, Beijing
100190, People's Republic of China}

\date{\today}

\begin{abstract}
Shubnikov-de Haas (SdH) oscillations   under a dc current bias are experimentally studied on a Hall bar sample of single layer graphene. In dc resistance, the bias current shows the common damping effect on the SdH oscillations and the effect can be well accounted for by an elevated electron temperature that is found to be linearly dependent on  the current bias. In differential resistance,  a novel phase inversion of the SdH oscillations has been observed  with increasing dc bias, namely we observe the oscillation maxima develop into minima and vice versa.  Moreover, it is found that the onset biasing current, at which a SdH extremum is about to invert,  is linearly dependent on the magnetic field of the SdH extrema. These observations are quantitatively explained with the help of a general SdH  formula.
\end{abstract}
\maketitle 

The effect of a dc current bias on  the nonlinear response of two-dimensional electron systems (2DES) 
in a classically strong magnetic field is a subject of current interest \cite{JQ_Zhang}.  In conventional 2DES,  current bias induced effects have been widely studied, in the context of  the breakdown of quantum Hall effect \cite{Eber, Cage},
and of some recently discovered nonlinear phenomena such as the Zener-tunneling oscillations \cite{Yang} and zero differential states \cite{Bykov}. Nevertheless, similar studies on 2DES with a relativistic-like linear energy dispersion, as recently realized in single layer graphene \cite{Novoselov,Zhang}, are less reported.

In this paper, we report on our experimental study on the influence of a relatively small dc bias 
on the magnetotransport of a single layer graphene. In the bias regime we explored (with current density up to 20 A/m), we find the magnetoresistance at lower field ($B<2$ T) has negligible dependence on dc bias, while the Shubnikov-de Haas (SdH) oscillations, occurring at higher fields, are obviously damped by increasing bias current. We show that the damping of the SdH oscillations can be well accounted for by an elevated electron temperature that is found to be linearly dependent on  the bias current.  

Our most important findings, however, are from the \emph{differential} resistance measurements, where a phase inversion of the SdH oscillations is observed with increasing the bias current. We observe the onset biasing current, at which a SdH maxima (minima) is about to invert to a minima (maxima),  is linearly dependent on the magnetic field of the SdH extrema. These novel observations are quantitatively explained by taking into account the nonlinear response of the SdH, due to elevated electron temperatures by the biasing current. 

Data presented in this paper were measured on a  lithographically defined Hall bar device of a single layer graphene,
as shown in Fig.~1(a). The single layer graphene was mechanically exfoliated \cite{Novoselov,Zhang} from Kish graphite onto degenerately doped silicon substrate with a $300$-nm thermal oxide SiO$_2$.  The Hall bar pattern was defined by electron-beam lithography (EBL) and oxygen plasma etching, with PMMA as a resist. Ohmic electrodes were defined by a second EBL, and by the subsequent 50nm-Pd deposition and lift-off processes.  

Transport measurements were performed on a PPMS system (Quantum Design) which can provide a base temperature of $2$ K and a magnetic field up to $14$ T. The sample was in-situ annealed for an hour to degas the sample surface before cooling down. The carrier density of the sample was tuned by a gate voltage $V_g$ applied to the Silicon substrate. The differential resistance, $r=\partial V/\partial I$, was measured with standard, low frequency ($30.9$ Hz) lock-in technique in the presence of both a small ($100$ nA) ac excitation current and a dc bias current, $I_{dc}$; while the dc resistance, $R=V/I$,  was measured  by a dc voltage meter in the presence of the dc bias current alone. 

The graphene sheet is identified to be single layer by the observation of half-integer quantum Hall plateaus, 
together with corresponding minima in magnetoresistance $R_{xx}$, as shown in Fig. 1b and Fig. 1c. The sample mobility is  generally higher than $8,000$  cm$^2$/Vs within the experimental window  $\vert V_g\vert \leqslant 40$ V. 
The Dirac point, $V_g\sim 1.5$ V, is found to be very close to zero gate voltage, which indicates the sample is clean. 
Strong SdH oscillations and their very good symmetry about zero magnetic field, as shown in Fig. 1c, imply high homogeneity of the sample.
\begin{figure}
\includegraphics{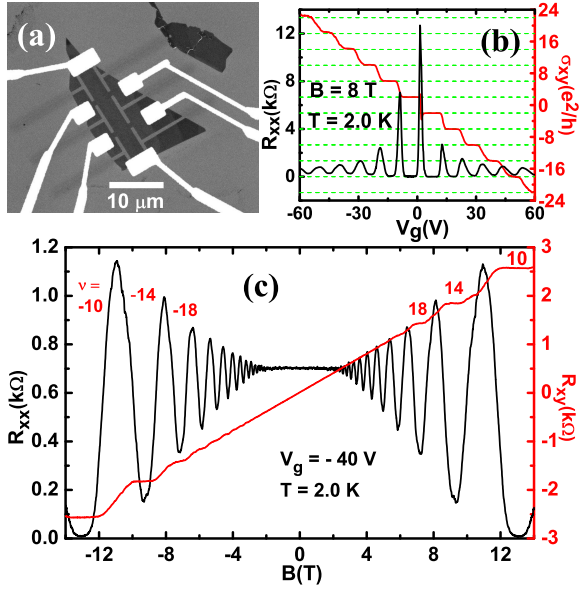}
\caption{(color online). (a) A SEM image of the graphene Hall bar device. The Hall bar  (black color) was defined by electron-beam lithography and oxygen plasma etching, with six Pd electrodes (white color). The width of the bar is about 2.8  $\mu$m, and the distance between voltage contacts along the same side is about 7  $\mu$m.  (b) The longitudinal magnetoresistance $R_{xx}$ and Hall conductivity $\sigma_{xy}$ against the gate voltage at fixed magnetic field $B=8$ T. The half-integer quantum Hall plateaus at $\sigma_{xy}=\nu e^2/h$ with $\nu=4(N+1/2), N=\pm0, \pm1, \pm2$,\dots , are hallmarks of a single layer graphene. (c) The magnetoresistance measured at a fixed gate voltage $V_g=-40$ V.}  
\end{figure}

In our sample, the most observable effects of a dc bias current are on the SdH oscillations. Typical  experimental traces 
are shown in Fig.2, which were measured at $T= 2.0$ K and with a fixed gate voltage  $V_g=-40$ V. 
As shown in Fig. 2, the magnetoresistance is nearly flat at lower field ($B<2$ T) and has negligible dependence on dc bias, while the SdH oscillations, occurring at higher fields, are obviously dampened by dc biasing. 

In dc resistance, the data shown in Fig. 2 (a) resemble clearly those of  temperature dependence measurements shown in Fig. 3 (a), implying an electron heating effect of the dc bias commonly observed on the magnetotransport of a 2DES. However, except for the amplitude damping, the \emph{differential} resistance  shown in Fig. 2 (b) manifests a novel feature that the SdH oscillation extrema are inverted with increasing the bias current. Moreover, it is found that the onset biasing current ($I_{inv}$), at which a SdH maxima (minima) is about to invert to a minima (maxima),  is linearly dependent on the magnetic field of the SdH extrema ($B_{ex}$), with a slope $\beta=4.2\ \mu$A/T, as shown in the inset of Fig. 2 (b).

\begin{figure}
\includegraphics{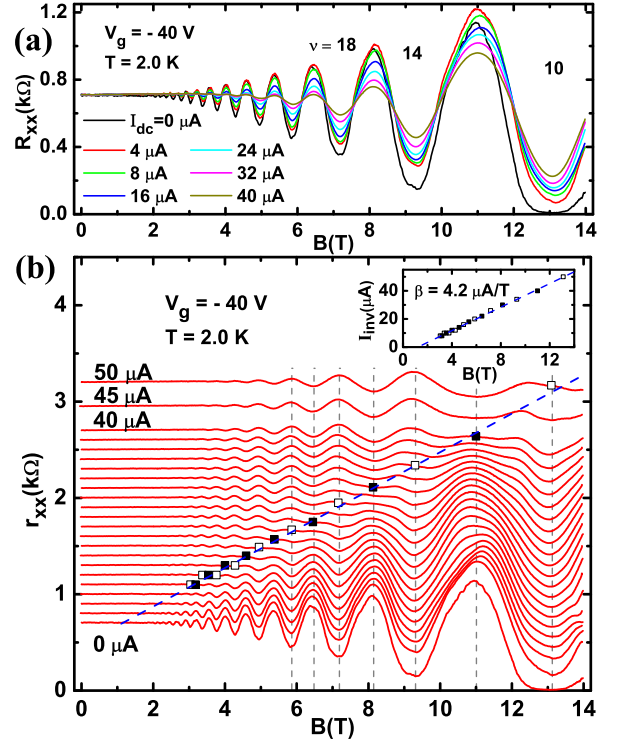}
\caption{ (color online). The dc bias dependence of magnetoresistance measured  at $V_g=-40$ V and T= 2.0 K.  (a)  Traces of dc resistance at selected bias currents. 
(b) Traces of  differential resistance  at various dc bias currents: from bottom to top,   $I_{dc}=0, 2.0, 4.0, \dots, 40.0\ \mu$A, in $2.0\ \mu$A steps for the lower 21 traces, and  $I_{dc}=45.0, 50.0\ \mu$A for the upper two traces, respectively.  The traces are shifted vertically for clarity. In differential resistance, the extrema of the SdH oscillations invert whenever the bias current is sufficiently large;  the onset traces of  the inversion are marked by open and filled squares in Fig. 2 (b), for selected SdH maxima and minima, respectively. The inset of Fig. 2(b) reveals that the onset bias current is  linearly dependent on the magnetic field of the SdH extrema, with a slope $\beta=4.2\ \mu$A/T.}
\end{figure}

. 
In the regime of SdH oscillations, the magnetoresistance of a 2DES can be wrote in a general form regardless of its energy dispersion\cite{Ando, Isihara, Coleridge,Gusynin}
 \begin{eqnarray}
R_{xx} = R_0\left[1+ \lambda \sum_{s=1}^{\infty} D(sX) \exp(-\frac{s\pi}{\omega_c \tau}) \right.  \nonumber\\
\left. \cos \left(s\frac{\hbar S_F}{eB}-s\pi+s\phi_0\right)\right],
 \end{eqnarray}
where $\lambda$ is a constant prefactor, $S_F=\pi k_F^2$ is the area enclosed by the Fermi circle, 
$\omega_c=eB/m^*$ is cyclotron frequency, $\tau$ is the lifetime of the carrier, and D(sX) is the temperature damping factor
\begin{eqnarray}
D(sX) =  \frac{sX}{\sinh (sX)}=\frac{s\ 2\pi^2 k_B T/\hbar\omega _c}{\sinh (s\ 2\pi^2 k_B T/\hbar\omega _c)}.
 \end{eqnarray}

In Eq. (1), $\phi_0$ accounts for the Berry phase of the 2DES,  with $\phi_0=0$ for conventional 2DES and $\phi_0=\pi$ for single layer graphene. Due to its linear energy dispersion $\epsilon(k)=v_F\hbar k$, the effective mass of a single layer graphene is dependent on the carrier density: 
\begin{equation}
m_c^*=\hbar k_F/v_F=(\hbar/v_F)\sqrt{\pi n_s}.
 \end{equation}
 
From Eq. (1), the amplitude of the SdH oscillations at each extremum ($B_{ex}$), neglecting higher harmonics to the first order, is given by  
\begin{equation}
A_{ex}=\lambda D(X) \exp(-\frac{\pi}{\omega_c \tau}).
\end{equation}
At sufficient high temperature such that $2\pi^2 k_B T/\hbar\omega _c>1$,  a linear relation on temperature for the quantity
\begin{eqnarray}
F(A_{ex},T)&\equiv&B_{ex}\ln \left( \frac{\hbar e}{8\pi^2 k_B m_c^*}\frac{B_{ex}}{T}A_{ex}\right)\nonumber\\
 &=& - \frac{2\pi ^2 k_Bm^*}{\hbar e}T+B_{ex}\ln(\frac{\lambda m^*}{2m_c^*})-\frac{\pi m^*}{e\tau }
\end{eqnarray}
follows, which can be used to extract the effective mass $m^*$, with  the theoretic mass $m_c^*$ calculated by Eq. (3).

Figure 3 (b) shows the plot $F(A_{ex},T)$ vs. $T$ for several SdH extrema of the traces presented in Fig.~3~(a). It is evident that data from different extrema collapse on the same line at temperature $T \ge 10 $ K, with a  slope corresponding to the calculated effective mass $m^*=m_c^{*}=(\hbar/v_F)\sqrt{\pi n_s}=0.0332m_e$, where the carrier density $n_s=3.16\times10^{12}/cm^2$ is obtained from the measured SdH period and $v_F=1.1\times10^6$ m/s adopted from literatures \cite{Zhang,Deacon}. This excellent agreement testifies the validity of Eq. (1) to describe the SdH oscillations in our graphene sample. 

\begin{figure}
\includegraphics{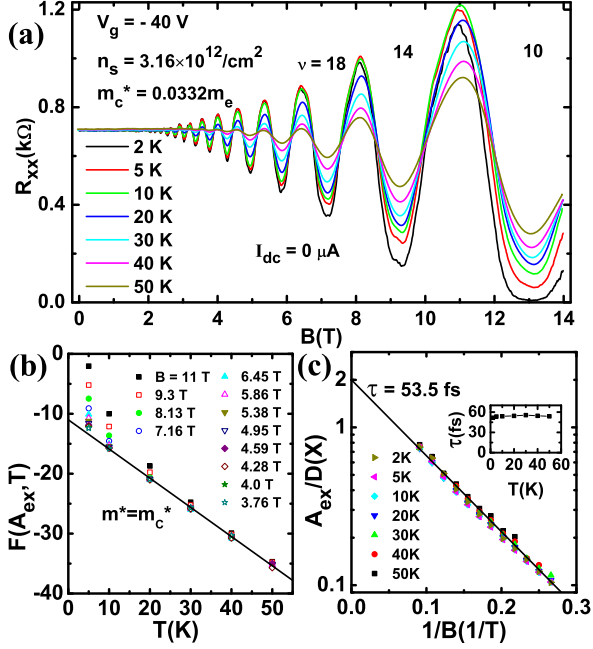}
\caption{ (color online). (a)  Temperature dependence of the  magnetoresistance measured at $V_g=-40$ V and zero bias.  (b) The $F(A_{ex},T)$ vs. T plot for various SdH extrema as labeled in the graph. 
The slope of the solid line corresponds to $m^*=m_c^*$, where $m_c^*$ is calculated by Eq.~(3). (c)  The  $A_{ex}/D(X)$ vs. $1/B_{ex}$ plot that reveals the life time of the carrier. The solid line  corresponds to $\tau =53.5$ fs. Note the vertical intercept of the line at $1/B=0$ indicates $\lambda=2.0$. The inset reveals that the lifetime is nearly constant in the temperature range.}
\end{figure}

With the effective mass known, Eq.~(4) suggests that the lifetime can be extracted from the slope of a $log(A_{ex}/D(X))$ vs. $1/B_{ex}$ plot. Such plots for the data shown in Fig. 3 (a) are presented in Fig. 3 (c). The inset of Fig. 3 (c)  indicates that the lifetime, $\tau \approx 54$ fs,  is nearly constant up to $T=50$ K, which is consistent with the observation of temperature-independent  resistance at low magnetic fields.

An interesting result from Fig.~3~(c) is that a prefactor $\lambda=2.0$ is obtained for the graphene sample, which is different from that of conventional 2DES where  $\lambda=4$ is theoretically predicted \cite{Isihara} and experimentally confirmed\cite{ Coleridge, Coleridge_1991}. 
There is theoretical implication \cite{Gusynin} that $\lambda =2$ is intrinsic to the Dirac fermions in graphene, however, in addition to our work, more experiments are demanding to fully test this point.

As previously mentioned, the resemblance between the data shown in Fig. 2 (a) and those shown in Fig. 3 (a) implies an electron heating effect of the dc bias on the SdH oscillations. To be more quantitative, the electron temperature, $T_e$,  can be extracted  by fitting  the experimental traces with  the SdH formula. In particular, with a constant lifetime in the studied regime,  Eq. (4) gives
\begin{eqnarray}
\frac{A_{ex}(T_e)}{ A_{ex}(T_0)} =\frac{\sinh (2\pi^2 k_B T_0/\hbar\omega _c)}{\sinh (2\pi^2 k_B T_e/\hbar\omega _c)}\frac{T_e}{T_0},
\end{eqnarray}
where  $A_{ex}(T_e)$ and $A_{ex}(T_0)$ are the amplitude of  a SdH extremum at $B=B_{ex}$, measured with or without a bias current, respectively, at the same base temperature $T=T_0$.  
In Fig. 4 we plot the electron temperature $T_e$, extracted via  the one parameter fitting to Eq. (6),  against $I_{dc}$. The results clearly confirm that $T_e \propto I_{dc}$,  with a slope $\alpha=1.07 $ K/$\mu$A. 

The linear dependence of  $T_e$ on $I_{dc}$ indicates that the energy loss of the electron system, 
$P=P_{joule}\propto I_{dc}^2 \propto T_e^2$,
implicating an dominant energy dissipation by the diffusion of the hot electrons into cold  electrodes, rather than by the emission of phonons \cite{Viljas}. Assuming simply the Wiedemann-Franz law, $\kappa = \mathcal{L} \sigma T_e$, between the thermal and  electrical conductivities,  the electron temperature can be estimated from the heat balance between the loss by electron diffusion and the joule heating $\nabla (\kappa\nabla T_e) = P_{joule}$ \cite{Viljas}.  And further assuming a quasi-one-dimensional solution along the Hall bar,  the electron temperature in the middle of the Hall bar is roughly $T_m \approx R_0/(2\sqrt{\mathcal{L}})I_{dc}$,
where $\mathcal{L}=\pi^2k_B^2/(3e^2)$ is the Lorenz number, and $R_0$ is the resistance of the Hall bar at zero magnetic field.Therefore, we can estimate an average electron temperature $T_e=T_m/2=\alpha I_{dc}$, with
\begin{equation}
 \alpha \approx \frac{R_0}{4\sqrt{\mathcal{L}}} = \frac{\sqrt{3}e}{4\pi k_B}R_0. 
\end{equation}
Taking the experimental value  $R_0\approx 700 \Omega$, we estimate $\alpha \sim 1.12 K/\mu$A for the data given in Fig. 2 (a), which agrees surprisingly well with the experimental value $\alpha \sim 1.07 K/\mu$A as obtained in Fig. 4. 

\begin{figure}
\includegraphics{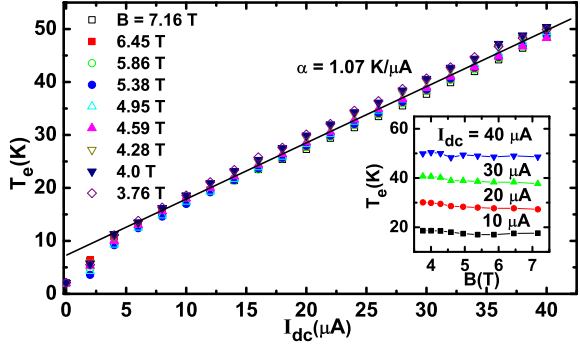}
\caption{ (color online). The electron temperatures, extracted from amplitudes of the SdH oscillations, as function of bias current. A linear relation with a slope $\alpha = 1.07\ K/\mu$A is found. The inset shows that, at a given bias current,  the electron temperature is roughly constant in the regime of SdH oscillations. }
\end{figure}

Having demonstrated the validity of Eq.~(1) for describing the SdH oscillations in the single layer graphene, and established that the effect of a bias current can be taken into account by an effective electron temperature $T_e$, now we are ready to focus on the differential resistance that is given by 
\begin{eqnarray}
r_{xx}\equiv \left(\frac{\partial V}{\partial I}\right)_{I_{dc}}= \frac{\partial (IR_{xx})}{\partial I}=R_{xx}+I_{dc}\frac{\partial R_{xx}}{\partial I}
\end{eqnarray}
where $R_{xx}$ is given by Eq. (1)  with $T=T_e(I_{dc})$. 
In the experimental regime, we have found  that $R_{0}(T)$ and $\tau$ are near constant with respect to the temperature or bias current, it follows 
\begin{eqnarray}
r_{xx}=R_{0}\left[1+\Lambda\cos \left( \hbar S_F/eB-\pi+\phi_0\right)\right],
\end{eqnarray}
where higher harmonics of the oscillatory terms have been neglected, and  the oscillation amplitude is
\begin{eqnarray}
\Lambda=\lambda\left(D(X_e)+I_{dc}\frac{\partial D(X_e)}{\partial T_e}\frac{\partial T_e}{\partial I_{dc}}\right)\exp(-\frac{\pi}{\omega_c \tau}), 
\end{eqnarray}
with $X_e= 2\pi^2 k_B T_e/\hbar\omega _c$. 

The second term in the bracket of right-hand side of  Eq. (10) is proportional to $I_{dc}$, but its sign is negative, opposite to the first term, 
because ${\partial D(X_e)}/{\partial T_e} < 0$, and normally we should have ${\partial T_e}/{\partial I_{dc}} >0$.  As a result, when the bias current is sufficiently large, the SdH amplitude of the differential resistance can become negative, giving rise to a inversion of oscillation extrema. The onset of the inversion occurs at 
\begin{eqnarray}
D(X_e)+I_{dc}\frac{\partial D(X_e)}{\partial T_e}\frac{\partial T_e}{\partial I_{dc}}=0.
\end{eqnarray}
In our sample, the electron temperature is linear dependent on bias current, such that the solution of Eq. (11) satisfies $X_e=1.915$, i.e., 
\begin{eqnarray}
k_BT_e/\hbar\omega_c=0.097,
\end{eqnarray}
thus we have the onset current for phase inversion 
\begin{eqnarray}
I_{inv}\approx T_e/\alpha =\beta B;   \text{with  } \beta= 0.097\hbar e/(k_B m^*\alpha), \ 
\end{eqnarray}
which explains well the observed relation as shown in Fig. 2 (b). Moreover, substitute the observed coefficient $\alpha=1.07$ K/$\mu$A  and the effective mass $m^*=0.0332m_e$ into Eq. (13), we get a coefficient $\beta=3.67\ \mu$A/T , which reasonably agrees with the value $\beta=4.2\ \mu$A/T determined from the experimental data. 

From the above analysis, we emphasize that the dc-bias-induced inversion of SdH oscillations is unique to the differential resistance measurements, unlike that of magneto-intersubband oscillations where the inversion originates in the dc resistance, as recently discovered in double quantum wells \cite{Bykov_2008, Mamani}.  It is evident that this phenomenon in differential resistance is generic in 2DES, regardless of their energy dispersion. 

We notice that similar dc-bias-induced inversion of SdH oscillations has been observed in conventional 2DES of high mobilities  \cite{Kalmanovitz, Studenikin}, where it is believed that the phenomenon 
cannot be simply described by an eleviated electron temperature, rather a nonuniform spectral diffusion has to be taken into account \cite{JQ_Zhang}. Our work indicates that, at least for 2DES  in the lower mobility regime,  the observed phase inversion of  SdH oscillations  can be well accounted for by an eleviated electron temperature.
 
 In summary,  we have studied the influence of a dc bias on the magnetoresistance of a single layer graphene.   
In dc resistance, electron temperatures extracted from the amplitude of SdH oscillations manifest an  linear dependence on  the  bias current, implicating a dominant heat dissipation mechanism via electron diffusion. In differential resistance,  a novel phase inversion of the SdH oscillations has been observed,  with an onset biasing current that is proportional to the magnetic field. 

\begin{acknowledgments}
We thank S. K. Su and H. F. Yang for experimental assistance. This work was supported by the NSFC (Grant No. 10874220), and by the Main  Direction Program of Knowledge Innovation of CAS (Grant No. KJCX2-YW-W30). 
\end{acknowledgments}

\end{document}